\documentclass[aps,prl,showpacs,twocolumn,floats]{revtex4}
\usepackage{amssymb}

\usepackage{psfig}
\usepackage{dcolumn}
\usepackage{bm}

\begin{document}

\bibliographystyle{prsty}

\title{Ignition of superconducting vortices by acoustic standing waves}
\author{Jaroslav Albert and E. M. Chudnovsky}
\affiliation{\mbox{Department of Physics and Astronomy,} \\
\mbox{Lehman College, City University of New York,} \\ \mbox{250
Bedford Park Boulevard West, Bronx, New York 10468-1589, U.S.A.}
\\ {\today}}
\begin{abstract}
Nucleation of vortices in a superconductor below the first critical
field can be assisted by transverse sound in the GHz frequency
range. Vortices will enter and exist the superconductor at the
frequency of the sound. We compute the threshold parameters of the
sound and show that this effect is within experimental reach.
\end{abstract}


\maketitle

A superconducting cylinder rotated at an angular velocity ${\bf
\Omega}$ about its symmetry axis develops a magnetic moment ${\bf M}
= -(mc/2\pi e){\bf \Omega}$, where $m$ and $e$ are bare electron
mass and charge, and $c$ is the speed of light. This effect
predicted by London \cite{London} has been subsequently tested in
experiment and proved with an accuracy to many significant figures.
It is a consequence of a more general gyromagnetic effect predicted
by Barnett \cite{Barnett}: ``A substance which is magnetic must
become magnetized by a sort of molecular gyroscopic motion on
receiving an angular velocity''. Barnet effect is, in its turn, a
consequence of the Larmor theorem \cite{Larmor}: In the rotating
frame of reference the action of the rotation on charged particles
is equivalent to the action of the magnetic field ${\bf H}_{\Omega}
= {\bf \Omega}/\gamma$ \cite{Larmor}, where $\gamma$ is the
gyromagnetic ratio. For electron's orbital motion $\gamma = e/(2mc)
\approx 0.9 \times 10^7\,$(Gauss)$^{-1}$ s$^{-1}$. Thus, in
practice, the fictitious field in the reference frame of a rotating
macroscopic cylinder can hardly exceed a fraction of a milligauss.
This would be well below the lower critical field ${\bf H}_{c1}$
when the temperature of the superconductor is not too close to
$T_c$. Due to the Meissner effect \cite{Meissner} (considered in the
frame of the rotating cylinder) such a field would be expelled from
the bulk of the cylinder by a superconducting current induced at the
surface. Writing ${\bf B} = {\bf H}_{\Omega} + 4 \pi {\bf M} = 0$
for the total field in the bulk, one obtains the London's magnetic
moment, ${\bf M} = -{\bf H}_{\Omega}/(4\pi) = -(mc/2\pi e){\bf
\Omega}$. Due to the symmetry of the problem it is the same in the
rotating and laboratory frames.

In this Letter we would like to take this problem a little further
and look at the consequence of an angular velocity well beyond the
experimental limit. In particular, we are interested in the
rotational velocity of a magnitude that would generate a fictitious
magnetic field that exceeds ${\bf H}_{c1}$. If Larmor's theorem
still holds, than it must be the case that a superconducting vortex
enters the bulk of the cylinder. This would require the angular
velocity to be of order $10^9$s$^{-1}$, clearly surpassing the
feasible experimental value for a mechanical rotation. While this
scenario is merely a thought experiment we will use it as a
motivation to study the effect of local rotations generated in a
superconductor by high frequency ultrasound. Interaction of sound
with vortices has been studied in the past \cite{Haneda,Dom,Sonin}.
Radiation of phonons by supersonic vortices \cite{Ivlev,BC}, phonon
contribution to the vortex mass \cite{Coffey,Duan,EC-Kuklov-PRL},
and decoherence of flux qubits by phonons
\cite{EC-Kuklov-PRB,Albert-EC-PRB} have been investigated. In this
Letter we are addressing a completely different problem --
possibility of the nucleation of a vortex by sound.

Within continuous elastic theory, local deformations are described
by the displacement vector field, ${\bf u}({\bf r},t)$. We will be
interested in the effect of transverse sound waves. Such waves
create shear deformations of the crystal lattice, such that
\begin{equation}\label{udot}
\nabla\cdot{\bf u}=0.
\end{equation}
In the long-wave limit they do not affect the density of the ionic
lattice but result in a local rotation at an angular velocity
\cite{LL}
\begin{equation}\label{omega}
{\bf \Omega}({\bf r},t)=\frac{1}{2}\nabla\times\dot{\bf u}\,.
\end{equation}
The frequency of ultrasound achievable in experiment with, e.g.,
surface acoustic waves can easily be in the ballpark of
$f\sim10^{10}$s$^{-1}$ \cite{Santos}. According to Eq.\
(\ref{omega}) a sound of such frequency and amplitude of a few
nanometers can provide $\Omega \sim 10^9$s$^{-1}$ that can generate
fictitious magnetic fields above ${\bf H}_{c1}$. For practical
purposes, it may be convenient to loosen the restriction on the
frequency and amplitude of ultrasound by applying an external
magnetic field ${\bf H}$ near, but less than, ${\bf H}_{c1}$. We
shall see that within one percent of ${\bf H}_{c1}$, vortices can be
ignited by the ultrasound in the GHz frequency range.

For a vortex to enter a superconductor, the Gibbs free energy of the
system must be lowered. We compute the extra free energy due to the
vortex and determine the condition at which it becomes negative. It
should be noted that the system under consideration is dynamical,
and therefore is not at a thermodynamical equilibrium. However, we
are interested in the free energy of the Cooper pairs which can
adjust to the changes of state in a time scale orders of magnitude
shorter than the period of the sound. This time scale is
proportional to the relaxation time $\tau$ of the cooper pairs, i.e.
$\tau\sim10^{-12}$s. As mentioned before, the period of the sound
$T=1/f$ will be always greater than $10^{-10}$s. Under these
conditions, our system is adiabatic and the thermodynamic
equilibrium can be safely established. The calculation that follows
is similar to the conventional calculation of ${\bf H}_{c1}$. The
presence of ${\bf \Omega}$, however, introduces a new feature into
this calculation so we will follow it all the way through to show
how the sound enters the problem.

It is convenient to calculate the extra free energy in terms of the
magnetic field and its spacial derivatives. The electric field
produced by the time derivatives will be neglected. The kinetic
energy of the superfluid is $\frac{1}{2}n_sm{\bf v}^2$ where $n_s$
is the number density of the superconducting electrons and
\begin{equation}\label{SFvel}
{\bf
v}=\frac{e^{\ast}}{m^{\ast}c}\left(\frac{\hbar}{e^{\ast}}\nabla\varphi-{\bf
A}\right)
\end{equation}
is the velocity of the cooper pairs with $\varphi$ and ${\bf A}$
being the phase of the superfluid wavefunction and the magnetic
vector potential, respectively. The stared quantities represent the
effective mass and charge of cooper pairs. We will take them to be
$m^{\ast}=2m$ and $e^{\ast}=2e$. The normal electrons experience
viscous forces as they move relative to the nuclei contributing zero
average normal current. The ionic charge per unit volume consisting
of the nuclei and the normal electrons is therefore exactly opposite
to that of the cooper pairs. The total current is then
\begin{equation}\label{current}
{\bf j}= en_s({\bf v}-\dot{\bf u}),
\end{equation}
where $\dot{\bf u}$ is the velocity of ions. Eq.\ (\ref{current})
reflects the fact that the electric current corresponds to the
motion of electrons relative to ions. It is invariant with respect
to the motion of the reference frame. With Eq. (\ref{SFvel}) in mind
we can write the gauge invariant current in terms of $\varphi$ and
${\bf u}$ as
\begin{equation}\label{GagInvJ}
{\bf j}=\frac{n_se\hbar}{2m}\left(\nabla\varphi-\frac{2e}{\hbar}{\bf
A}_{eff}\right),
\end{equation}
where
\begin{equation}\label{Aeff}
{\bf A}_{eff}={\bf A}+\frac{mc}{e}\dot{\bf u}
\end{equation}
is the effective vector potential felt by the electrons in the
rotating frame of the ions \cite{EC-Kuklov-PRL}. In terms of the
total current ${\bf j}$, the kinetic energy of the superconducting
electrons may be expressed in the form
\begin{equation}\label{SCKE}
KE_e=\int d^3r\frac{n_sm}{2}\left(\frac{1}{n_se}{\bf j}+\dot{\bf
u}\right)^2.
\end{equation}
The energy of the sound is
\begin{equation}\label{EneSound}
E_s=\int d^3r\frac{1}{2}(\rho_0\dot{\bf
u}^2-\lambda_{iklm}u_{ik}u_{lm})
\end{equation}
in which $\rho_0$ is the combined mass density of ions and normal
electrons, $\lambda_{iklm}$ is the tensor of elastic coefficients
and $u_{ik}=\frac{1}{2} (\partial_iu_k+\partial_ku_i)$ is the strain
tensor. Using Maxwell's equation $\nabla\times{\bf B}=(4\pi/c){\bf
j}$ and combining Eqs. (\ref{SCKE}) and (\ref{EneSound}) the
expression for the total Gibbs free energy yields
\begin{eqnarray}\label{freeEne}
G&=&{\cal F}_0+\frac{1}{8\pi}\int d^3r\left[{\bf
 B}^2+\frac{\lambda^2}{f(r)}(\nabla\times{\bf B})^2\right]\nonumber\\
&+&\frac{1}{4\pi}\int d^3r\frac{mc}{e}\dot{\bf
u}\cdot(\nabla\times{\bf B})-
\frac{1}{4\pi}\int d^3r{\bf H}\cdot{\bf B}\nonumber\\
&+&\int d^3r\frac{1}{2}\left[\rho\dot{\bf
u}^2-\lambda_{iklm}u_{ik}u_{lm}\right],
\end{eqnarray}
Here, ${\cal F}_0$ is the free energy in the absence of currents,
fields, and sound, $\lambda=\sqrt{mc^2/4\pi n_se^2}$ is the London
penetration depth, $f(r)=(|\psi|/|\psi_\infty|)^2$ in which $\psi$
is the complex order parameter and $|\psi_\infty|=\sqrt{n_s/2}$ is
the order parameter in the absence of gradients and fields, and
$\rho = \rho_0+n_sm$ is the total mass density of the
superconductor. The fourth term can be recognized as the interaction
of the external magnetic field with the magnetization. It is this
term that is responsible for the nucleation of vortices in the
absence of sound when $H\geq H_{c1}$.

Before we can calculate the free energy of Eq. (\ref{freeEne}) we
must first work out the magnetic field. This can be done by
replacing the current in the Maxwell's equation $\nabla\times{\bf
B}=(4\pi/c){\bf j}$ with Eq. (\ref{GagInvJ}) and defining a gauge
invariant vector potential ${\bf Q}={\bf A}-(\hbar
c/2e)\nabla\varphi$, so that we obtain the following equation:
\begin{equation}\label{Q}
\lambda^2\nabla\times(\nabla\times{\bf Q})+f(r){\bf
Q}=-\frac{mc}{e}f(r)\dot{\bf u}.
\end{equation}
For $\nabla\varphi=0$ (${\bf Q}={\bf A}$) Eq. (\ref{Q}) becomes
equivalent to the London's equation with a source. When a vortex
enters a superconductor the phase must be quantized according to the
condition $\oint\nabla\varphi\cdot d{\bf l}=2\pi$. For certainty we
consider a transverse standing sound wave having one node at the
center of a superconducting slab of thickness $d$ large compared to
the coherence length $\xi$. The external field is applied parallel
to the slab, see Fig. 1. In this case $\lambda_s=2d$. Generalization
to standing waves with many nodes is straightforward. If the field
is close to ${\bf H}_{c1}$, a vortex will periodically enter and
exit the slab. The boundary condition on the current is ${\bf
J}_\perp \cdot {\bf n}=0$, where ${\bf n}$ is the direction of the
surface. If the thickness of the slab $d$ is of order or less then
$\lambda$, this boundary will distort the cylindrical symmetry of
the vortex. We can satisfy the boundary condition by placing image
vortices of alternating sign a distance $d$ apart on the outside of
the slab. The equation for the magnetic field, in the region $r>\xi$
where $|\psi|=1$, can then be written in two parts, namely ${\bf
B}={\bf B}_0+{\bf B}_v$, such that the first term satisfies
\begin{equation}\label{B0}
\lambda^2\nabla\times(\nabla\times{\bf B}_0)+{\bf
B}_0=-\frac{2mc}{e}{\bf \Omega},
\end{equation}
while the second is a solution of
\begin{equation}\label{Bv}
\lambda^2\nabla\times(\nabla\times{\bf B}_v)+{\bf B}_v=\Phi_0{\bf
e}_z\sum_{n=-\infty}^{\infty}(-1)^n\delta({\bf r}+nd{\bf e}_x),
\end{equation}
where $\Phi_0=hc/2e$ is the flux quantum. Notice that Eqs.
(\ref{B0}) and (\ref{Bv}) can be obtained by taking a curl of Eq.
(\ref{Q}) with the account of the vortex cores represented by the
delta functions.

Since we are interested in standing sound waves we can choose the
displacement vector ${\bf u}$ to be
\begin{equation}\label{u}
{\bf u}({\bf r},t)=u_0\sin(kx)\sin(\omega t){\bf e}_y.
\end{equation}
The quantity $k=\omega/v=2\pi/\lambda_s=\pi /d$ is the wave number
with $\lambda_s$ and $v$ being the wavelength and the speed of sound
respectively. It is easy to see from Eq.\ (\ref{omega}) that
$\Omega$ is maximum at the nodes.
\begin{figure}
\unitlength1cm
\begin{picture}(18,5.5)
\centerline{\psfig{file=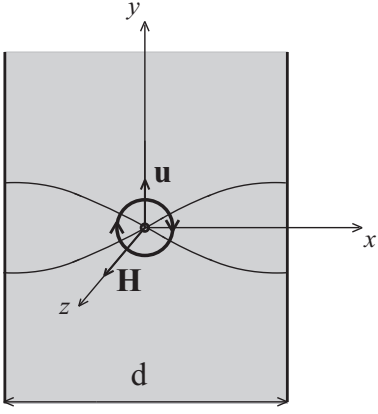,width=5.7cm}}
\end{picture}
\caption{Standing wave in a slab with one node at the center. The
vortex is generated at the node where $\Omega$ is maximum.}
\label{fig1}
\end{figure}
The corresponding solutions to Eqs. (\ref{B0}) and (\ref{Bv}) with
the boundary condition ${\bf B} = {\bf H}$ at $x=\pm d/2$ are
\begin{eqnarray}
{\bf B}_0(x) & = & {\bf B}_M + {\bf B}_s \label{B0Sol} \\  {\bf
B}_v({\bf r}) & = & \sum_{n=-\infty}^{\infty}(-1)^n{\bf b}({\bf
r}_n) \label{BvSol}
\end{eqnarray}
where
\begin{eqnarray}
& &{\bf B}_M  =  2{\bf H}\,\frac{\sinh(d/2\lambda)}
{\sinh(d/\lambda)}\cosh\left(\frac{x}{\lambda}\right) \label{BM}\\
& &{\bf B}_s = -\frac{2mc}{e}\frac{\bf \Omega}{1+k^2\lambda^2}
 \label{Bs}  \\
& &{\bf b}({\bf r}_n)=\frac{\Phi_0}{2\pi\lambda^2}K_0(|{\bf
r}+nd{\bf e}_x|/\lambda){\bf e}_z\label{b}\,,
\end{eqnarray}
and $K_0$ is a zeroth-order Hankel function of imaginary argument.
The first term in Eq. (\ref{B0Sol}) is the Meissner field while the
second is due to the sound.

Let us now integrate by parts the third term in Eq. (\ref{freeEne})
and insert ${\bf B}={\bf B}_0+{\bf B}_v$. By doing so we obtain
\begin{eqnarray}\label{freeEne3}
& & G = G_0+\Delta E+\frac{1}{4\pi}\int dr^3{\bf B}_v\cdot\left[
\frac{2mc}{e}{\bf \Omega}+{\bf B}_s\right]\nonumber\\
& &+ \frac{1}{4\pi}\int d^3r\frac{\lambda^2}{f(r)}(\nabla\times{\bf
B}_v)\cdot(\nabla\times{\bf B}_s)
\end{eqnarray}
where $G_0$ is the Gibbs free energy without a vortex and
\begin{eqnarray}\label{DeltaE}
\Delta E&=&\frac{1}{4\pi}\int
d^3r\frac{\lambda^2}{f(r)}(\nabla\times{\bf
B}_M)\cdot(\nabla\times{\bf
B}_v)\nonumber\\
&+&\frac{1}{8\pi}\int d^3r\left[{\bf
B}_v^2+\frac{\lambda^2}{f(r)}(\nabla\times{\bf
B}_v)^2\right]\nonumber\\
&-&\frac{1}{4\pi}\int dr^3{\bf H}\cdot[{\bf B}_v+{\bf B}_M]
\end{eqnarray}
is the vortex energy. One can simplify the volume integrals in Eq.
(\ref{freeEne3}) by separating the integration over the core from
the integration over the volume outside of the core. When the latter
is integrated by parts, the integrals outside the core cancel, and
the free energy in Eq. (\ref{freeEne3}) with the help of Eq.
(\ref{B0}) becomes
\begin{equation}
\Delta{\cal F}=\Delta{\cal F}_1+\Delta{\cal F}_2+\Delta{\cal
F}_3+\Delta E\,,
\end{equation}
where
\begin{eqnarray}
& &\Delta{\cal F}_1=\frac{1}{4\pi}\int_c dr^3 {\bf
B}_v\cdot\left[\frac{2mc}{e}{\bf \Omega}+{\bf
 B}_s\right]\label{DeltaF1}\\
& &\Delta{\cal F}_2=\frac{\lambda^2}{4\pi}\oint_c{\bf
B}_v\times(\nabla\times{\bf B}_s)\cdot d{\bf s} \label{DeltaF2} \\
& &\Delta{\cal F}_3=\frac{1}{4\pi}\int_c
d^3r\frac{\lambda^2}{f(r)}(\nabla\times{\bf
B}_v)\cdot(\nabla\times{\bf
B}_s).\label{DeltaF3} \nonumber\\
\end{eqnarray}
The subscript $c$ indicates an integration over the core. The
surface integral in Eq.\ (\ref{DeltaF2}) is over the boundary of the
normal core.  Near the vortex core $f(r)=(r/a)^2$, where
$a\approx\xi$. It is straightforward to check that in the limit
$r\rightarrow 0$ the exact solution to Eq. (\ref{Q}) for the vector
potential ${\bf A}_s({\bf r})$ is
\begin{equation}\label{As}
{\bf A}_s({\bf r})=-\frac{2mc}{e}\Omega_0x{\bf e}_y\,.
\end{equation}
Then the magnetic field ${\bf B}_s=\nabla\times{\bf A}_s({\bf r})$
generated by the sound at the center of the core is
\begin{equation}\label{Bs2}
{\bf B}_s({\bf r})=-\frac{2mc}{e}\Omega_0{\bf e}_z,
\end{equation}
where
\begin{equation}\label{DeltaNout}
\Omega_0=\frac{1}{2}u_0k\omega=\frac{\pi}{2}\frac{u_0}{d}\omega.
\end{equation}
It can be shown that near the vortex core, $\nabla\times B_v\propto
r^4$ and $\nabla\times B_s\propto r^5$. The expression under the
integral in Eq. (\ref{DeltaF3}) is therefore proportional to $r^8$
near the center of the core and to $rK_1(r/\lambda)$ at $r\gtrsim
\xi$. Thus, the integral in Eq. (\ref{DeltaF3}) falls off very
rapidly inside the core and can be neglected.

The case of $k\lambda \geq 1$ is rather involved as it requires
explicit knowledge of the structure of the vortex core. For
$k\lambda\ll1$ Eq. (\ref{Bs}) provides that ${\bf
B}_s\cong-(2mc/e){\bf \Omega}$ in all regions of space, so that
$\Delta{\cal F}_1\rightarrow 0$. In this limit the Meissner field
${\bf B}_{M}$ and the fields due to images can be neglected. The
total interaction energy per unit length of the vortex acquires the
simplest form at $\kappa=\lambda/\xi\gg1$:
\begin{equation}\label{DelF2Solution}
\frac{\Delta{\cal F}_2}{L} = -\frac{mc}{2\pi}\Omega_0\Phi_0
\left(\frac{k\lambda}{\kappa}\right)^2\ln\kappa\,,
\end{equation}
where $L$ is the dimension of the slab in the z-direction.

If one excludes small contribution from the vortex core in Eq.
(\ref{DeltaE}), then the integration by parts yields
\begin{equation}\label{DeltaE2}
\frac{\Delta E}{L}=\frac{\lambda^2}{8\pi}\oint {\bf
B}_v\times(\nabla\times{\bf B}_v)\cdot d{\bf s} -\frac{1}{4\pi}\int
d^3r{\bf H}\cdot{\bf B}_v\,.
\end{equation}\\
This approximation is good if $\lambda$ and $d$ are large compared
to the coherence length $\xi$. Then the vortex energy per unit
length is
\begin{equation}\label{DeltaESol}
\frac{\Delta E}{L} = \frac{\Phi_0^2}{({4\pi\lambda})^2}\ln\kappa -
\frac{\Phi_0H}{4\pi}\,.
\end{equation}
The first term in this expression is the self-energy of the vortex,
while the second term is the energy of the interaction of the flux
quantum with the external field.

The condition for the nucleation of the vortex, $\Delta{\cal F}_2 +
\Delta E = 0$, yields
\begin{equation}\label{ConFin}
\frac{2mc}{e}\Omega_0\left(\frac{k\lambda}{\kappa}\right)^2\ln\kappa=
{\epsilon}H_{c1}\,,
\end{equation}
where
\begin{equation}
\epsilon = 1 - \frac{H}{H_{c1}}
\end{equation}
and $H_{c1}=\Phi_0\ln \kappa/(4\pi\lambda^2)$ is the first critical
field that follows from Eq.\ (\ref{DeltaESol}) at $\Delta E = 0$.
Substituting Eq. (\ref{DeltaNout}) into Eq. (\ref{ConFin}), one
finds the conditions on the frequency $f$ and amplitude $u_0$ of the
sound needed to nucleate a vortex in the geometry shown in Fig.
\ref{fig1}:
\begin{equation}\label{ConFin2}
f = \frac{v}{2d}\,, \qquad u_0=\frac{\epsilon}{4} \left(\frac{d}{\pi
\lambda}\right)^4 \frac{\hbar\kappa^2}{mv}\,.
\end{equation}
While the last formula was derived under the conditions $\pi\xi <
\pi\lambda \ll d$, our numerical analysis shows that it holds even
for $d \sim \pi\lambda$ at $\kappa \gg 1$ and is true by order of
magnitude for $\kappa \sim 1$. In this case the expression for
$H_{c1}$ carries the signature of the surface barrier \cite{Bean}:
$H_{c1}=\beta \Phi_0\ln \kappa/(4\pi\lambda^2)$, where
\begin{equation}
\beta = \frac{1-{2({\ln \kappa})^{-1}\sum_1^\infty (-1)^n
K_0(dn/\lambda)}}{1 -
2{\sinh(d/2\lambda)}[{\sinh(d/\lambda)}]^{-1}}\,.
\end{equation}

For the speed of the transverse sound $v \sim 3\times 10^5$cm/s, in
a slab of thickness $d \sim \pi\lambda \sim 6\times 10^{-5}$cm and
$\kappa \sim 2$, with $H$ within one percent of $H_{c1}$, one gets
from Eq.\ (\ref{ConFin2}) $f \sim 3$GHz and $u_0 \sim 0.2$nm. These
are accessible values of frequency and amplitude of ultrasound.

As ${\bf \Omega}$ changes its sign every half a period of the sound,
vortices are periodically attracted and repelled by the standing
acoustic wave in Fig. \ref{fig1}. Periodic entering and expulsion of
vortices should result in the elevated attenuation of the ultrasound
and in the ac voltage across the slab at the sound frequency. In a
different experiment one can assist vortices to enter or exit the
superconductor with the help of the surface acoustic waves (SAW).
Like in the problem with a slab, local rotation of the crystal
produced by the SAW may assist nucleation of the vortex at the field
just below $H_{c1}$.

In conclusion, we have demonstrated that nucleation of a vortex in a
superconductor can be assisted by ultrasound. In the presence of a
standing sound wave, vortices will periodically enter and exit the
superconductor. The required amplitude and frequency of ultrasound
are within experimental reach.

We thank Lev Bulaevskii and Carlos Calero for useful discussions.
This work has been supported by the Department of Energy through
Grant No. DE-FG02-93ER45487.

\end{document}